\documentclass[a4paper]{jpconf}
\usepackage{graphicx,amsmath,amssymb}

\def\be{\begin{equation}}
\def\ee{\end{equation}}
\def\bea{\begin{eqnarray}}
\def\eea{\end{eqnarray}}

\def\gev{\, {\rm GeV}}
\def\mev{\, {\rm MeV}}

\newcommand{\sigmaSD}{\sigma_{\rm SD}}

\newcommand{\pb}{\rm pb}

\newcommand{\m}{\rm m}
\newcommand{\cm}{\rm cm}
\newcommand{\s}{\rm s}
\newcommand{\sr}{\rm sr}
\newcommand{\yr}{\rm yr}
\newcommand{\kT}{\rm kT}

\begin{document}
\title{New Dark Matter Search Strategies at DUNE }

\author{Carsten Rott}

\address{Department of Physics, Sungkyunkwan University, 2066 Seobu-ro, Suwon 440-746, Korea}

\ead{rott@skku.edu}

\author{Seongjin In}

\address{Department of Physics, Sungkyunkwan University, 2066 Seobu-ro, Suwon 440-746, Korea}

\ead{seongjin.in@gmail.com}

\author{Jason Kumar}

\address{Department of Physics \& Astronomy, University of Hawai'i, 2505 Correa Road, Honolulu, HI 96822, U.S.A.}

\ead{jkumar@hawaii.edu}

\author{David Yaylali}

\address{Department of Physics, University of Arizona, 1118 E. Fourth Street, Tucson, AZ 85721, U.S.A.}

\ead{yaylali@email.arizona.edu}

\begin{abstract}
If dark matter annihilates to light quarks in the core of the Sun, then a flux
of 236 MeV neutrinos will be produced from the decay of stopped kaons.  We consider
strategies for DUNE to not only observe such a signal, but to determine the direction
of the neutrino from the hadronic recoil.  We show that this novel strategy can provide a
better handle on systematic uncertainties associated with dark matter searches.
\end{abstract}

\section{Introduction}

One of the most important strategies for the indirect detection of dark matter is to search for
neutrinos produced by dark matter annihilation inside the core of the Sun~\cite{Silk:1985ax,Press:1985ug,Krauss:1985ks}.
Typically, one  focuses on searching for a continuous spectrum of high-energy neutrinos, produced by
dark matter annihilation, above the atmospheric neutrino background.  In this proceedings contribution, we will describe
a different possibility~\cite{Rott:2015nma,Rott:2016mzs}; we will find that there is a broad class of models
in which dark matter annihilation in the Sun produces sub-GeV monoenergetic neutrinos, and find that
neutrino detectors with excellent energy resolution (such as DUNE) can resolve this type of line signal.
We will also find that detectors with good angular resolution can resolve the direction of the incoming
neutrino, despite its low energy.  This result provides a new dark matter search strategy with a better handle on
systematic uncertainties.

It has long been appreciated that, if dark matter annihilates to light quark pairs ($\bar u u$, $\bar d d$, $\bar s s$)
in the Sun,  then
hadronization processes will produce light, long-lived mesons which will interact with the dense
solar medium and will tend to stop before they decay.  As a result, the secondary neutrinos produced by hadron
decay will be very soft.  Such low-energy neutrinos would not only have a relatively small scattering cross section
with the target nuclei in a neutrino detector, but would also compete against an atmospheric neutrino background
which increases at low energies.  As such, these light quark annihilation channels have long been ignored when
studying dark matter annihilation in the Sun.

But it was pointed out in~\cite{Rott:2012qb,Bernal:2012qh} that the interactions of these light mesons with the
dense nuclear medium tend to produce hadronic cascades with many light mesons, which mostly decay at rest.
One essentially trades a hard
neutrino spectrum for a softer spectrum, but a larger flux.
These works focused on the continuum $\bar \nu_e$ flux produced by stopped meson
decay and subsequent matter and vacuum oscillation effects.
It was shown that water Cherenkov detectors can be
sensitive to this soft neutrino flux through inverse beta decay, an ideal channel due to its large cross section,
potentially providing evidence for dark matter interactions.

We instead consider the monoenergetic neutrino produced by the stopped meson decay process $K^+ \rightarrow \mu^+ \nu_\mu$,
which has a branching fraction of $\sim 64\%$.
This 236 MeV monoenergetic neutrino essentially constitutes a line signal which can be resolved by a neutrino
detector with good energy resolution, such as DUNE~\cite{Kumar:2015nja,Rott:2015nma}.  Interestingly, at neutrino detectors with good
angular resolution and particle identification capabilities, one can also determine if the neutrinos arrive from the
direction of the Sun.  This seems counter-intuitive; when sub-GeV neutrinos have a charged-current interaction in
the detector, they produce charged leptons which are distributed roughly isotropically.  But for a 236 MeV neutrino, many
charged-current events with an argon nucleus target will result in the ejection of proton which will be forward-directed.  This
provides a new type of dark matter search strategy which dramatically reduces uncertainties, and for which DUNE is well-suited.

We will see that signal sensitivity of this type of search is limited, when compared to dark matter searches at direct detection
experiments (such as PICO) or using water Cherenkov detectors (which benefit from a much larger exposure).  But where this
search strategy really shines is as a complementary strategy which can dramatically reduce systematic uncertainties.

\section{236 MeV Neutrinos at DUNE}

If a dark matter particle scatters off a solar nucleus, it can lose enough kinetic energy to be gravitationally
captured, in which case it will orbit the Sun and continue to scatter until eventually settling into the solar
core.  The rate at which dark matter is captured ($\Gamma_C$) is determined by the dark matter mass ($m_X$) and scattering cross
section ($\sigma$),
as well as by some astrophysics parameters (such as the ambient dark matter density) and by the composition of the
Sun~\cite{Gould:1987ir,Gould:1991hx,Wikstrom:2009kw,Danninger:2014xza} (uncertainties in the capture rate are discussed
in~\cite{Choi:2013eda}).
We will focus on the case of spin-dependent scattering ($\sigma = \sigmaSD$), since this is the case in which the
sensitivity of neutrino detectors is most competitive with that of direct detection experiments.

The dark matter which collects in the core of the Sun can then annihilate to produce Standard Model particles, whose
subsequent decays produce neutrinos which can be probed by neutrino detectors.  For the range of model parameters in which
we are interested, the dark matter is in equilibrium~\cite{Kumar:2012uh}, implying that the total annihilation rate ($\Gamma_A$) is related to
the capture rate by $\Gamma_A = (1/2) \Gamma_C$.  Given a choice of the dark matter annihilation channel, a bound on the
neutrino flux from the Sun arising from dark matter annihilation corresponds to a bound on $\Gamma_A$, which in turn constrains
the $(\sigmaSD, m_X)$ parameter space.  To determine the sensitivity of such a search, one need only
determine the number of signal events expected from dark matter annihilation, as well as the number of background
events expected from atmospheric neutrinos.

The number of events is given by
\bea
N_{S,B}^{e, \mu} &=&
\left[f_{S,B}  \eta_{S,B}^{e, \mu} \int_{E_0-\Delta E/2}^{E_0+\Delta E/2} dE \int d\Omega \, \frac{d^2 \Phi_{S,B}^{e, \mu} }{dE d\Omega} \right]
\times \left[ T  A_{\rm eff}^{e, \mu}  \right]   ,
\label{eq:numEvents}
\eea
where $d^2 \Phi_{S,B}^{e, \mu} / dE d\Omega$ is the differential neutrino flux at the detector,
$E_0 = 236$ MeV is the energy of the monoenergetic neutrino arising from stopped kaon decay,
$\Delta E$ is the width of the energy bin over which one integrates,
$T$ is
the live-time, $A_{\rm eff}$ is the effective area for a charged-current interaction, and $\eta$ is the
efficiency for events to pass selection cuts.  The subscripts $S$ and $B$ refer to signal and background
events, while the superscripts $e$ and $\mu$ refer to electron or muon events, respectively.  $f$ is the fraction
of neutrino events which fall within the energy bin of width $\Delta E$; we may take $f_B =1$, since the atmospheric
neutrino background is relatively smooth, while we take $f_S =0.68$, since we will assume $\Delta E = \epsilon E$,
where $\epsilon$ is the fractional energy resolution.

The flux of 236 MeV neutrinos arising from dark matter annihilation in the Sun is
\bea
\frac{d^2 \Phi_S^{e, \mu}}{dE d\Omega} &=& \frac{(\Gamma_C / 2)F^{e,\mu} }{4\pi r_{\oplus}^2 } \left( 0.64 \times \frac{2m_X}{m_K} r_{K}(m_X) \right)
\delta (E-E_0) \delta(\Omega),
\eea
where $r_{\oplus} \sim 1.5 \times 10^{11}\m$ is the Earth-Sun distance, $m_K$ is the mass of $K^+$,
and the factor of $0.64$ is the branching fraction for the decay $K^+ \rightarrow \mu^+ \nu_\mu$.
The $\delta$-functions arise because the neutrinos are monoenergetic, and point back to the core
of the Sun.

The dark matter capture rate may be expressed as
$\Gamma_C = C_0^{SD} (m_X) \times \sigmaSD^p $, where $\sigmaSD^p$ is the
spin-dependent dark matter-proton scattering
cross section and the coefficients $C_0^{SD} (m_X)$ can be found, for example, in~\cite{Gao:2011bq,Kumar:2012uh}.
$r_K (m_X)$ is the fraction of the dark matter mass which is converted into stopped $K^+$, and was determined in~\cite{Rott:2015nma}.
The $F^{e,\mu}$ are the fraction of the 236 MeV neutrinos which arrive at the detector as $\nu_{e,\mu}$, respectively,
after the effects of matter and vacuum oscillations; they can be determined from~\cite{Lehnert:2007fv}.

The atmospheric neutrino background fluxes at 236 MeV are given by~\cite{Battistoni:2005pd}
\bea
\frac{d^2 \Phi_B^e}{dE d\Omega} &\sim& 1.2 ~\m^{-2} \s^{-1} \sr^{-1} \mev^{-1} ,
\nonumber\\
\frac{d^2 \Phi_B^\mu}{dE d\Omega} &\sim& 2.3 ~\m^{-2} \s^{-1} \sr^{-1} \mev^{-1} .
\eea
As this background is fairly smooth, one finds
$N_B^{e,\mu} \propto (d^2 \Phi_B^{e, \mu} / dE d\Omega) \times (\epsilon E)$.

Finally, the effective area can be written as
\bea
A_{\rm eff}
&=&  (5.1 \times 10^{-10}\m^2 )
\left( \frac{\sigma_{\nu {\rm -Ar}}}{10^{-38}\cm^2 } \right) \left( \frac{M_{\rm target}}{34~\kT} \right) ,
\label{eq:effArea}
\eea
where $M_{\rm target}$ is the fiducial target mass, and $\sigma_{\nu {\rm -Ar}}$ is the neutrino-argon scattering cross section,
which can be determined with numerical tools such as \texttt{NuWro}~\cite{Golan:2012wx}.

We now have all of the pieces necessary to determine the signal-to-background ratio ($N_S / N_B$) and
signal significance ($N_S / \sqrt{N_B}$) obtainable for any neutrino detector.  They scale with the
detector parameters $A_{\rm eff}$, $T$, $\epsilon$,
and $\eta_{S,B}^{e,\mu}$ as
\bea
\frac{N_S}{N_B} \propto \frac{\eta_S}{\epsilon \eta_B}, \qquad
\frac{N_S}{\sqrt{N_B}} \propto \frac{\eta_S}{\sqrt{\eta_B}} \sqrt{\frac{T A_{\rm eff}}{\epsilon}} .
\eea
If we assume no selection cuts beyond the identification of a
charged lepton contained in the fiducial volume, then we may set $\eta_{S,B}^{e,\mu}=1$.  With the nominal choices
$\epsilon =0.1$, $T A_{\rm eff} = 34~\kT~\yr$, we find that one would expect about $\sim 50$ background $\nu_e$ events in
the energy bin centered at 236 MeV~\cite{Rott:2015nma}.  For dark matter with mass in the $4-10~\gev$ range, an observation consistent with
background only would allow the 90\% CL exclusion of models with $\sigmaSD^p \sim 2-4 \times 10^{-38}\cm^2$~\cite{Rott:2015nma}.  For such
models, the signal-to-background ratio would be $\sim 1/5$~\cite{Rott:2015nma}.

\section{Directionality}

Even for an exposure of $34~\kT~\yr$, one already has a significant number of background events.  One natural way to
reduce this background would be to preferentially select events in which the neutrino arrives from the direction of
the Sun.  If one could select such events, then one would dramatically improve the signal-to-background ratio, and
potentially improve signal significance as well.

Essentially, one would like $N_S / N_B > \delta N_B / N_B$, where $ \delta N_B / N_B$ is the fractional systematic
uncertainty in the background.  This would ensure that any persistent excess in the event rate is actually the result of
neutrinos arising from dark matter annihilation, and not merely the result of underestimating the background event
rate.  A directionality measurement helps with both sides of the inequality.  One can measure the background event
rate by counting events in which the neutrino arrives from a direction away from the Sun; this measurement effectively
reduces $\delta N_B / N_B$.  Conversely, one expects to increase the signal-to-background ratio $N_S / N_B$ by counting
events in which the neutrino arrives from the direction of the Sun.

Unfortunately, for sub-GeV neutrinos, one cannot determine the directionality of the neutrino from the direction of
the charged lepton, since the charged lepton is produced roughly isotropically.  However, the hadronic recoil is largely
forward peaked.  In particular, at this energy, many of the scattering events are quasi-elastic charged current (QECC) events
in which a single proton is ejected from the nucleus ($\nu_\ell + {}^{40}{\rm Ar} \rightarrow
\ell + p + {}^{39}{\rm Ar}$), and this proton will typically be ejected in the forward direction.  If this proton can
be identified and if its direction can be determined with sufficient precision, then one can cut on the direction of
a 236 MeV neutrino.

We determine the effect of these cuts by simulating $10^5$ $\nu_{e, \mu}$-argon scattering events using \texttt{NuWro}~\cite{Golan:2012wx}.
We can restrict ourselves to QECC interactions, since we have found that these interactions by far dominate the
events which pass the cuts we will impose.  But we will make no attempt to realistically model the detector response, but will
instead estimate detector efficiency based on thresholds given in the DUNE Conceptual Design Report~\cite{Acciarri:2015uup}.

Based on~\cite{Acciarri:2015uup}, we will assume that a charged lepton can be identified at DUNE if it has a kinetic energy of
$E_{\rm kin} > 30\mev$.  We will also assume that a proton can be identified if it has a kinetic energy of
$E_{\rm kin} > 50\mev$; we will refer to this as a ``tight" threshold.  But we will also consider the more optimistic possibility that
a proton can be identified at DUNE if it has a kinetic energy as small as $20\mev$, which will denote as a ``loose" threshold.
In all cases, we make the simplifying assumption that a particle can be identified with 100\% efficiency if its kinetic
energy is above the threshold, and is always missed if its kinetic energy is below the threshold.
We will thus select neutrino scattering events in which one can identify exactly one charged lepton and exactly one
nucleon, which is a proton, in the fiducial volume.  Note, we thus include multi-nucleon knockout events which satisfy the
selection criteria (only one identified proton).
The fraction of quasi-elastic neutrino scattering events satisfying these cuts will be denoted by
$\eta_{sel}$, and is the same for both signal and background events.  These selection efficiencies are listed for each
channel and threshold choice in Table~\ref{tab:SelectionCutEfficiency}.

\begin{table}[h]
\centering
\begin{tabular}{|c|c|c|}
  \hline
  particle ID & proton threshold  & selection efficiency ($\eta_{\rm sel}$) \\
  \hline
  electron & tight ($E_{\rm kin} > 50\mev$) & $0.43$ \\
  muon & tight ($E_{\rm kin} > 50\mev$) & $0.28$ \\
  \hline
  electron & loose ($E_{\rm kin} > 20\mev$) & $0.83$ \\
  muon & loose ($E_{\rm kin} > 20\mev$) & $0.75$ \\
  \hline
\end{tabular}
\caption{The fraction of QECC events which result in the production of a single charged lepton with
$E_{\rm kin} > 30\mev$ and a single ejected proton satisfying the listed cuts.
\label{tab:SelectionCutEfficiency}
}
\end{table}

\begin{figure}[h!]
  \centering
  \includegraphics*[width= 0.49 \textwidth]{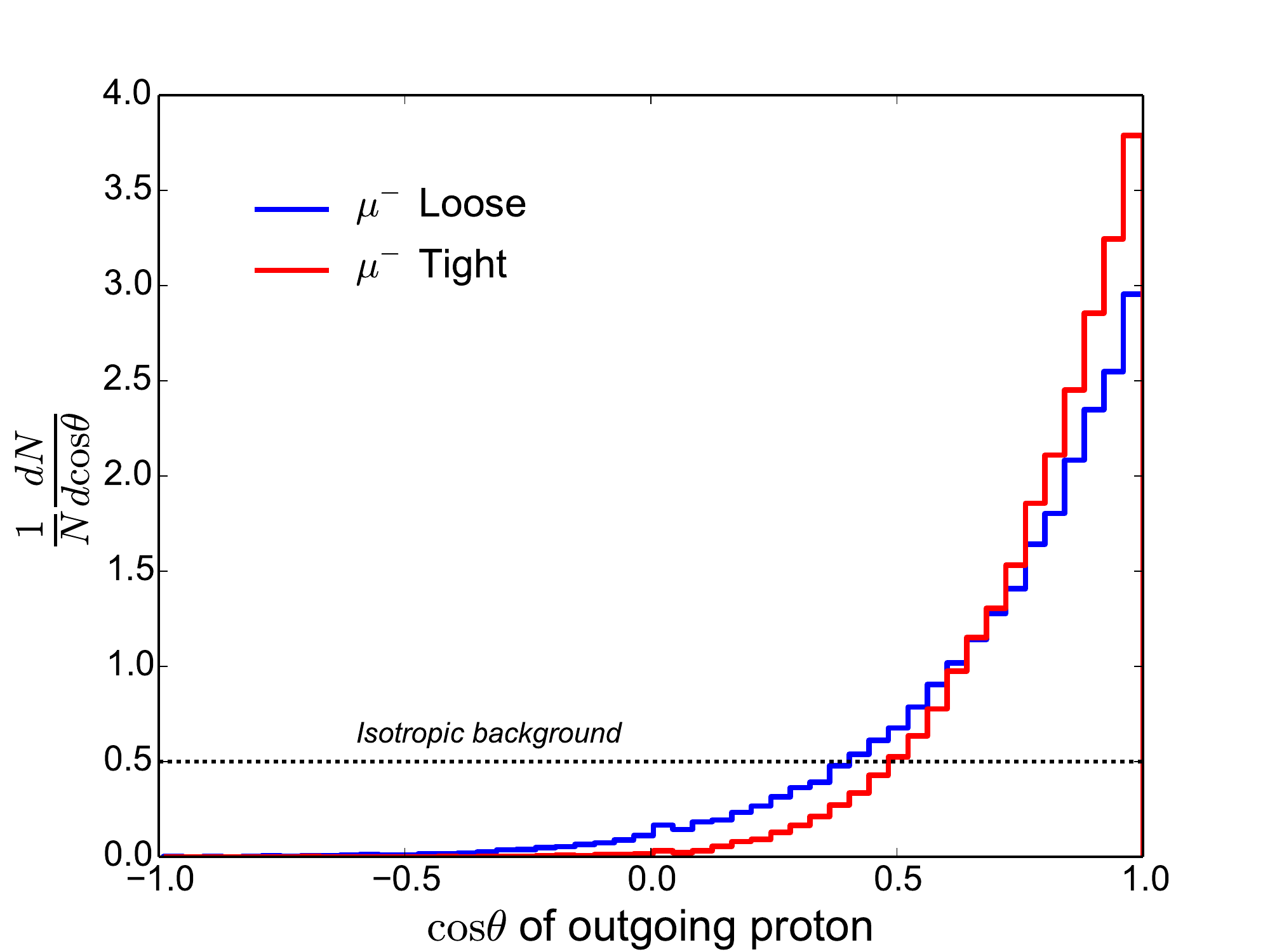}
  \includegraphics*[width= 0.465 \textwidth]{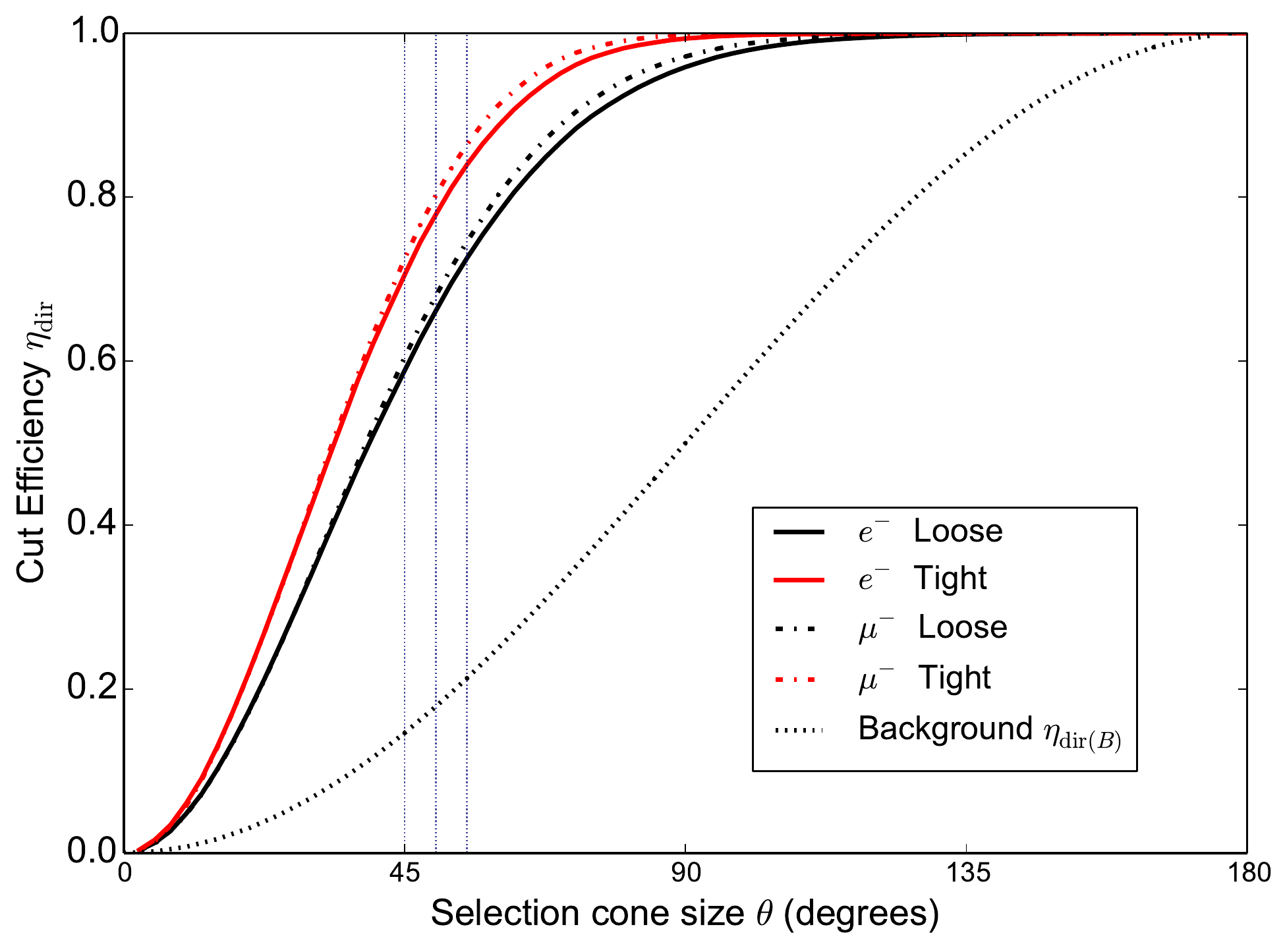}
  \caption{\label{fig:CosTheta} The left panel shows the number density of $\nu_{\mu} + {}^{40}{\rm Ar}$ events which pass selection cuts as a function of angle $\theta$ of the outgoing proton, showing that protons are typically ejected in the forward direction (the $\nu_e$ channel is
  essentially identical).
  The right panel shows the directionality cut efficiencies $\eta_{{\rm dir}(S)}^{e,\mu}$ for tight and loose signal events, as well as
  for background.
  The angle is measured from the direction vector of the incoming neutrino.  The small number of multi-nucleon knockout events which pass selection cuts are included in these distributions.
  The vertical lines (right panel) correspond to the angles which maximize the signal statistical significance.}
\end{figure}

In Figure~\ref{fig:CosTheta} (left panel), we plot the distribution of events passing the selection cuts as a function of the angle of the proton
with respect to the Sun,
while in the right panel we plot $\eta_{{\rm dir}}^{e,\mu}$, the fraction of events passing the selection cuts which lie within a cone of the Sun with a given
half-angle.  The angular distribution of background events is purely geometric.  The total efficiency for either signal or background events to
pass the event selection and directional cuts is given by $\eta_{S,B}^{e, \mu} = \eta_{{\rm sel}(S,B)}^{e,\mu} \times \eta_{{\rm dir}(S,B)}^{e,\mu}$.
For a given model parameter choice $(m_X, \sigmaSD^p)$ and detector exposure, the signal significance after the imposition of direction cuts is rescaled
by a factor $\eta_S / \sqrt{\eta_B}$, and the vertical lines in right panel of Figure~\ref{fig:CosTheta} denote the choices of cone half-angle which
maximize the improvement in signal significance for each of the channels and threshold choices.  Similarly,
the signal-to-background ratio is rescaled by a factor  $\eta_S / \eta_B$.
In Table~\ref{tab:Cuts}, we list
the cut angles, as well as the efficiencies and the enhancement to the signal-to-background ratio and signal significance, for each
channel and energy threshold.  Note that in all cases the cone half-angle is $\sim 45-55^\circ$; the angular resolution for an identified proton is
estimated to be $\sim 5^\circ$~\cite{Acciarri:2015uup}, which is negligible in comparison.

\begin{table}[h]
\centering
\begin{tabular}{|c|c|c|c|c|c|}
  \hline
  cut & half-angle & $\eta_S$ & $\eta_B$ & $S/B$ & sensitivity \\
   & & & & enhancement & enhancement \\
  \hline
  tight: electron & $45^\circ$ & $\eta_S^e = 0.30$ & $\eta_B^e = 0.06$ & 5.0 & 1.2 \\
  tight: muon & $50^\circ$ & $\eta_S^\mu = 0.23$ & $\eta_B^\mu = 0.05$ & 4.6 & 1.0 \\
  \hline
  loose: electron & $55^\circ$ & $\eta_S^e = 0.60$ & $\eta_B^e = 0.18$ & 3.3 & 1.4 \\
  loose: muon & $55^\circ$ & $\eta_S^\mu = 0.56$ & $\eta_B^\mu = 0.16$ & 3.5 & 1.4 \\
  \hline
\end{tabular}
\caption{ The cone half-angle, in the direction from the Sun, within which the ejected
proton track must lie for each of the listed cuts.
Also given is the total efficiency
of each set of cuts for signal ($\eta_S$) and background ($\eta_B$) events, as well
as the enhancement to the signal-to-background ratio obtained by applying each
set of cuts.
The last column gives the factor by which sensitivity is enhanced, for a fixed exposure, by
the application of the given directional cuts.
\label{tab:Cuts}}
\end{table}

\begin{figure}[ht]
\centering
\includegraphics[scale=0.60]{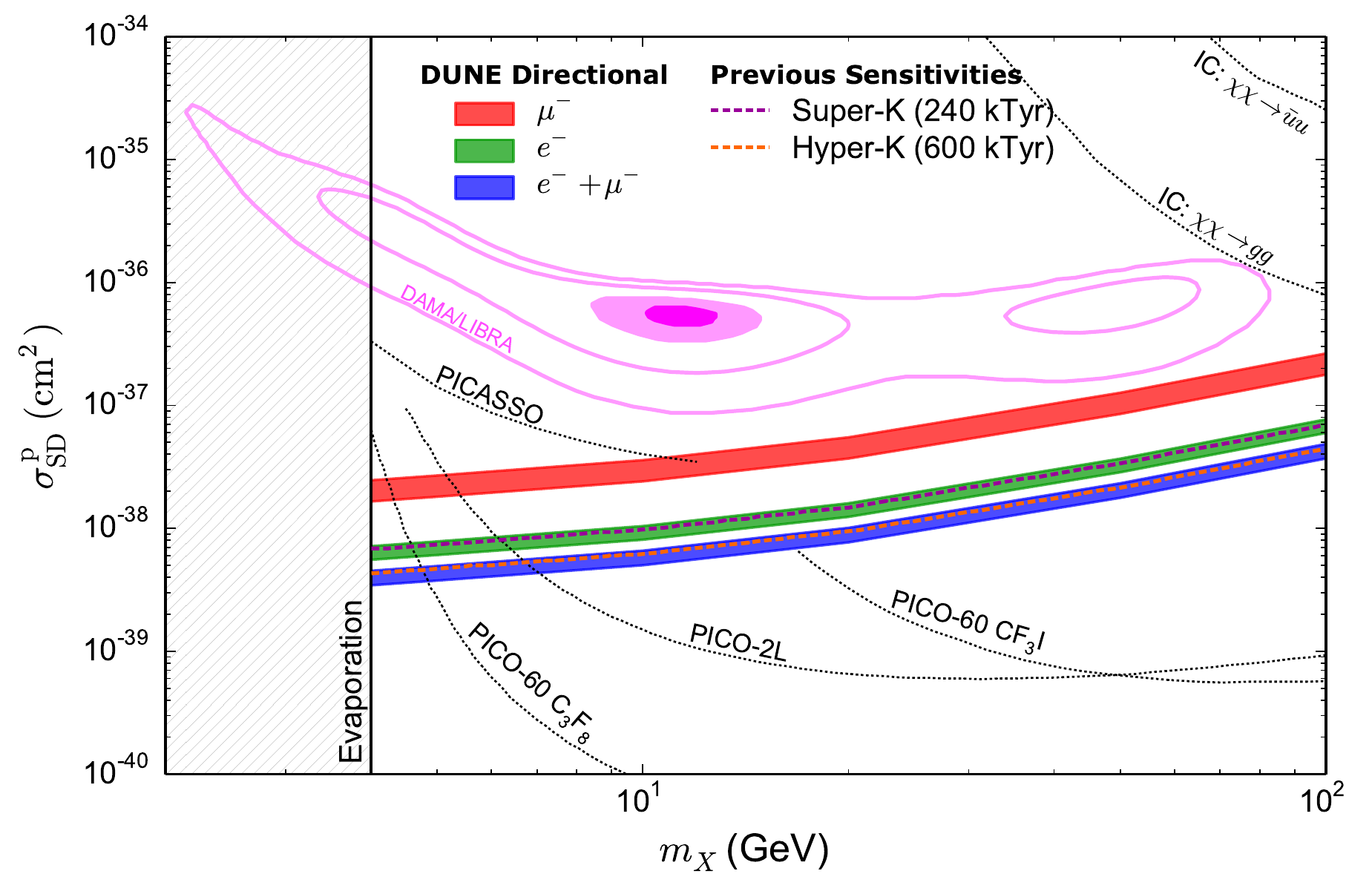}
\caption{Experimental sensitivity of DUNE at 90\%CL (with an exposure of 340 kT yr) using the directional search described in this work.  The bands span the sensitivity probed using the tight proton energy threshold requirement (upper edges) and the more optimistic loose threshold energy requirement (lower edges).  The blue band represents the combined sensitivities from the electron-only and muon-only analysis.
Also shown are previous sensitivities from DUNE~\cite{Acciarri:2015uup}, Super-K~\cite{Fukuda:2002uc}, and Hyper-K~\cite{Abe:2015zbg,Abe:2011ts} when directional information is not used, as found in~\cite{Rott:2015nma}.  We also show the exclusion regions based on IceCube data~\cite{Aartsen:2016exj,Aartsen:2012kia} computed with \texttt{nulike}~\cite{nulike,Aartsen:2016exj,Scott:2012mq}, on PICO-60~\cite{Amole:2015pla,Amole:2017dex} and on PICO-2L~\cite{Amole:2016pye},
and the region favored by DAMA/LIBRA (at $90\%/3\sigma/5\sigma/7\sigma$ CL)~\cite{Savage:2008er}.
\label{fig:ExclusionContour}
}
\end{figure}

In Figure~\ref{fig:ExclusionContour}, we plot the 90\% CL exclusion contours, assuming an exposure
of 340 kT yr, and assuming the number of observed events is consistent with the background-only
hypothesis.  DUNE's sensitivity decays rapidly for $m_X \lesssim 4\gev$, due to dark matter evaporation.

\section{Applications and Uncertainties}

As one can see from Figure~\ref{fig:ExclusionContour}, even a very large exposure at DUNE is only
competitive with direct detection experiments such as PICO in a very small mass range ($\sim 4-5~\gev$).
Even this range may shrink, if future iterations (PICO-40L, PICO-500) obtain even smaller recoil energy
thresholds.  Moreover, non-directional searches for dark matter annihilation in the Sun using water Cherenkov (WC)
detectors will always obtain a higher signal significance than DUNE, due to the much larger exposures which
are possible with WC detectors, compared to liquid argon time proejection chambers (LArTPCs).  Nevertheless
the ability to obtain directional information on the monoenergetic neutrino provides very good reasons
to perform this type of search at DUNE, by providing a distinct handle on astrophysical and detector
uncertainties.

The $\sim 5-10~\gev$ mass range is where PICO is rapidly losing sensitivity, and is thus very
sensitive to systematic uncertainties.   But neutrino search strategies are more robust against
detector and astrophysical uncertainties in this mass range, as indicated by the flatness of the
sensitivity curve.  In particular, whereas direct detection experiments search for low-mass dark matter
are very sensitive to the choice of dark matter velocity distribution, searches for dark matter annihilation
are much more robust against systematic uncertainties in the velocity profile~\cite{Choi:2013eda,Danninger:2014xza}.
Although searches using WC detectors are also robust against such astrophysical uncertainties, they
cannot determine the direction of the incoming  236 MeV neutrino, because the ejected proton will not have enough
kinetic energy to create a Cherenkov cone.  As such they suffer from detector-based systematic uncertainties which can be
largely eliminated at a LArTPC by comparing on-source and off-source event rates.  This unique handle on systematic uncertainties
implies that DUNE can play an important complementary role.

As one example of a scenario in which DUNE would play a vital role in the search for dark matter, consider the case
in which the solar system is currently in a local dark matter underdensity.  Since the event rate at a direct detection experiment
is proportional to the current dark matter density in the solar system, the sensitivity of such experiments could be
heavily suppressed.  But if the Sun is in equilibrium, then the dark matter annihilation rate in the Sun is determined
by the dark matter density averaged over the equilibration timescale, essentially smoothing out the effects of any
short-term variation in the dark matter density.

As another example, consider the scenario in which direct detection experiments detect dark matter in
$4-5~\gev$ mass range.  This is the exactly the mass range at which one would expect to see an asymmetric
dark matter candidate (see~\cite{Kumar:2013vba}, for example, for a review),
and one would naturally wish to explore this possibility by determining whether or not the dark matter
self-annihilated.  The observation of a monoenergetic 236 MeV neutrino signal pointing from the direction of the
Sun would be striking evidence of dark matter annihilation to light quarks, allowing one to distinguish symmetric
from anti-symmetric dark matter.  Moreover, for $\sigmaSD^p \sim {\cal O}(10^{-39}~\cm^2)$ and $m_X \sim 5~\gev$,
the Sun would still be in equilibrium even if the dark matter annihilation cross section were as
small as ${\cal O}(10^{-4})~\pb$~\cite{Kumar:2012uh}.  Even for such a small annihilation cross section, one would still see a striking
monoenergetic neutrino signature pointing from the direction of the Sun, with the equilibrium condition ensuring that
the small annihilation cross section is compensated by a large dark matter density.  It is difficult to see how such a
small annihilation cross section could be probed in any other way.

\section{Conclusions}

If dark matter annihilation in the core of the Sun produces light quarks, then one
expects a flux monoenergetic 236 MeV neutrinos from the Sun, arising from the decay of stopped $K^+$.  This
monoenergetic flux can be observed with neutrino detectors with excellent energy resolution, such as LArTPCs
like DUNE.  Moreover, one can also significantly reduce systematic uncertainties by selecting events in which
the neutrino arrives from the direction of the Sun, utilizing the fact that the ejected proton arising from
the QECC interaction is forward-directed.  This type of directional search for sub-GeV neutrinos is a unique
capability of detectors such as DUNE with excellent energy and angular resolution and particle identification.

Thus far, however, we have not been able to seriously model the detector response.  The biggest uncertainties lie
in the differential neutrino-nucleus scattering cross section~\cite{Akbar:2017dih}, which determines both the event rate and the
angular distribution.  As we have noted, one can reduce these uncertainties by comparing on-source and off-source event
rates.  But one can also directly calibrate the response of a DUNE by placing a LArTPC at a stopped pion experiment.
Any stopped pion experiment is also a stopped kaon experiment~\cite{Spitz:2012gp}, and there are plans in the works for calibrating LArTPCs
at such experiments.  For example, DAE$\delta$DALUS~\cite{Conrad:2010eu} is a planned stopped pion experiment to be placed at DUNE.  In addition,
CAPTAIN~\cite{Berns:2013usa} is a LArTPC which will be used for calibrating DUNE, and there are plans to place it at a stopped pion
experiment at the Spallation Neutron Source (SNS).

{\bf Acknowledgments}

We are grateful to Joshua Spitz, Robert Svoboda and Amanda Weinstein for useful discussions.
SI is supported by Global PH.D Fellowship Program through the National Research Foundation of Korea (NRF) funded by the Ministry of Education (NRF-2015H1A2A1032363).
The work of JK is supported in part by NSF CAREER grant PHY-1250573.
The work of CR is supported in part by
the Korea Neutrino Research Center which is established by the National Research Foundation of Korea (NRF) grant funded
by the Korea government (MSIP) (No. 2009-0083526) and Basic Science Research
Program NRF-2017R1A2B2003666.
The work of DY is supported
in part by DOE grant DE-FG02-13ER-41976 (DE-SC0009913).

\end{document}